\documentclass{webofc}
\usepackage[varg]{txfonts}   

\woctitle{?????????}
\begin{document}

\hspace{4.3in} \mbox{CERN-TH-2016-151}

\title{ Massless Interacting Scalar Fields  in de Sitter space}
%
%

\author{\firstname{Diana} \lastname{L\'opez Nacir}\inst{1}\fnsep\thanks{\email{diana.laura.lopez.nacir@cern.ch}} \and
        \firstname{Francisco D.} \lastname{Mazzitelli}\inst{2}\fnsep\thanks{\email{fdmazzi@cab.cnea.gov.ar}} \and
        \firstname{Leonardo G.} \lastname{Trombetta}\inst{2}\fnsep\thanks{\email{lgtrombetta@cab.cnea.gov.ar}}
}

\institute{Theoretical Physics Department, CERN, CH-1211 Gen\`eve 23, Switzerland.
\and
          Centro At\'omico Bariloche and Instituto Balseiro, CONICET, Comisi\'on Nacional de Energ\'\i a At\'omica,\\R8402AGP Bariloche, Argentina.
          }

\abstract{%
 We present a method to compute the two-point functions for an $O(N)$ scalar field model in de Sitter spacetime, avoiding 
 the well known infrared problems for massless  fields.  The method is based on  an exact treatment of the Euclidean zero modes and a perturbative one of the nonzero modes, and involves a partial resummation of the leading secular terms. This resummation,
 crucial to obtain a  decay of the correlation functions, is implemented along with a double expansion in an effective coupling constant $\sqrt\lambda$ and in $1/N$.   The results reduce to those known in the leading infrared approximation and coincide with the ones obtained directly in Lorentzian de Sitter spacetime in the large $N$ limit. The new method allows for a systematic calculation of higher order corrections both in  $\sqrt\lambda$ and in $1/N$.}

\maketitle

\section{Introduction}

The analysis of interacting quantum fields in de Sitter (dS)  spacetime is relevant to understand different aspects in cosmology, both in the early universe and in the present period of accelerated expansion.  Indeed, quantum effects could induce large corrections during inflation, including  non-Gaussianities in the cosmic microwave background. They could also produce nonnegligible contributions to the present cosmological constant, or could lead to instabilities of dS solutions to the semiclassical Einstein equations, in which the mean value of the stress tensor associated
to the quantum fields is used as a source. The case of light or massless fields is of particular importance, because it is in this case that perturbative calculations fail due to the infrared (IR) growth of the correlation functions.  

The case of a free scalar field already shows some peculiarities that occur in dS spacetime. For massive fields, it is possible to define a dS-invariant vacuum state (the so called Bunch-Davies vacuum). For this state, the Feynman propagator reads
\begin{equation}\label{Fulprop}
G_F^{(m)}(x,x')=\frac{ H^2\Gamma (\frac{3}{2}-\nu) \Gamma (\frac{3}{2}+\nu) }{(4\pi)^{2} }  \,_2F_1\left(\frac{3}{2}-\nu,\frac{3}{2}+\nu;2;1-\frac{r}{4}-i\epsilon\right),
\end{equation} where $\nu=\sqrt{{9}/{4}-{m^2}/{H^2}}$ and $r={[-(\eta-\eta')^2+|\vec{x}-\vec{x'}|^2]}/{\eta\eta'}$ (in comoving coordinates).
In the limit $r\to+\infty$, i.e. at large spatial separations or late times, the massive free propagator decays as
\begin{equation}
 G_{F}^{(m)}(r) \sim r^{-\frac{m^2}{3H^2}}.
\end{equation}
Expanding the propagator for masses  much smaller than the Hubble constant    $m^2\ll H^2$, the result behaves as $G_F^{(0)}\simeq3 H^4/(8\pi^2 m^2)$. There is a divergence for $m\to 0$ (which is not present in Minkowski spacetime),  and after subtracting this divergent term, the corresponding massless Feynman propagator 
 \begin{equation}
\hat{G}^{(0)}_F(r)= G^{(0)}_F(r)- \frac{3H^4}{8 \pi^2 m^2}\sim \log( r) \end{equation} 
shows a  logarithmic divergence as $r \to +\infty$. We see that the massless limit is not smooth: the behavior of the propagator is completely different in the IR. Moreover,
 for strictly massless fields there is no dS-invariant vacuum state \cite{Allen}.

When considering interacting fields (for instance a $\lambda\phi^4$ theory), the usual perturbative calculations involve
the above mentioned propagators. It can be shown that 
 a diagram with $L$ loops will be proportional to 
$(\lambda H^2/m^2)^L$ \cite{Burgess}. Therefore, the usual perturbative calculations break down for
light ($m^2\ll \lambda H^2$) or massless quantum fields. 

 There have been several attempts to cure the IR problem, all of them introducing some sort of nonperturbative approach.
For instance, in the so called stochastic approach \cite{Starobinsky} (which is  based on a classical treatment  of the  long  wavelength modes), nonperturbative results can be obtained at the leading IR order (i.e. at leading order in the coupling constant)  by  neglecting the interactions of the short wavelength modes.    In the context of QFT  in Lorentzian de Sitter spacetime, 
   the nonperturbative approaches are based on the    
   mean field (Hartree) approximation \cite{Hartree1},  the large $N$ expansion \cite{LargeN}, 
 the analysis of the Schwinger-Dyson equation using a physical momentum decomposition \cite{Gautier} or the exact renormalization group equation \cite{Guilleux}. A well known (nonperturbative) result  is that  interaction terms in the potential turn on a dynamical mass, and one can obtain finite and de Sitter invariant 
correlation functions  even for minimally coupled massless fields.
 However,  in none of these approaches, there is a clear systematic way of computing corrections beyond the leading IR approximation. 
The main inconvenients are of technical nature.
 
 An alternative and simpler approach emerges for quantum fields in Euclidean de Sitter space. As this is a compact space,  the modes of a 
 quantum field are discrete, and the origin of the IR problems can be traced back to the zero mode \cite{Rajaraman1}: the IR behavior 
 improves if the zero mode is treated exactly while the nonzero modes (UV modes in what follows) are treated
 perturbatively.  Using this idea, it has been shown that in the massless $\lambda\phi^4$ theory one obtains a dynamical squared mass for the field  that in the 
 leading IR limit is proportional to  $\sqrt{\lambda} H^2$.  Indeed, in the Euclidean space, this mass cures the IR problems. 
   Moreover, a systematic perturbative procedure for calculating  the $n$-point functions has been delineated   in Ref.~\cite{BenekeMoch}, where it was shown that, for massless fields, the effective coupling is $\sqrt\lambda$ instead of $\lambda$. It has been pointed out  that this procedure together with an analytical continuation  could  be used  to   cure  the IR problems also in the Lorentzian de Sitter spacetime, and  in particular to obtain $n$-point functions that respect the de Sitter symmetries.
 However,  so far explicit calculations have been restricted to obtaining corrections to the variance of the zero mode, which has no analog quantity in the Lorentzian de Sitter spacetime. 
The main concern here is to present  an explicit calculation of the inhomogeneous two-point functions of the scalar fields which,  after  the analytical continuation, lead to two-point functions  respecting  the de Sitter symmetries.
  
 In a recent work \cite{nos jhep}, we presented a generalization of the  approach of Ref.~\cite{BenekeMoch} to the case of $O(N)$ scalar field theory,  including
a detailed calculation of the corrections to the two-point functions up to second order in the 
 parameter $\sqrt \lambda$. We extended the nonperturbative treatment performing a resummation
of the leading IR secular terms, that produces the proper decay of the two-point functions at large distances.  When this resummation is combined with an expansion
 both in $\sqrt\lambda$ and $1/N$, it is possible to compute systematically
 the corrections coming from the interactions of both IR and UV sectors.  In what follows we will summarize our  findings in Ref.~\cite{nos jhep}, emphazising  the 
 main concepts and
 avoiding technical details.

\section{$O(N)$ model in Euclidean de Sitter space}

We will consider an $O(N)$-symmetric model with action
\begin{equation}
 S = \int d^d x \sqrt{g} \left[ \frac{1}{2} \phi_a \left( -\square + m^2 \right) \phi_a + \frac{\lambda}{8N} ( \phi_a \phi_a )^2 \right],
\end{equation}
where the fields $\phi_a$,  with $a=1,..,N$, are the components of an element of the adjoint representation of the $O(N)$ group, and the sum over repeated indices is implied. Euclidean de Sitter space is obtained from Lorentzian de Sitter space in global coordinates by performing an analytical continuation $t \to -i ( \tau - \pi/2H )$ and a compactification in imaginary time $\tau = \tau + 2\pi H^{-1}$. The resulting metric is that of a $d$-sphere of radius $H^{-1}$
\begin{equation}
 ds^2 = H^{-2} \left[ d \theta^2 + \sin(\theta)^2 d\Omega^2 \right].
\end{equation}
where $\theta = H\tau$. 
Due to the symmetries and compactness of this space, the field can be expanded in $d$-dimensional spherical harmonics
\begin{equation}
 \phi_a(x) = \sum_{\vec{L}} \phi_{\vec{L},a} Y_{\vec{L}}(x)\, .
\end{equation}

We generalize to the $O(N)$ model the method 
developed in \cite{Rajaraman1} and \cite{BenekeMoch} for a single field, for which the constant zero modes $\phi_{0a}$ play a crucial role. We split the fields as  $\phi_a(x) =  \phi_{0a} + \hat{\phi}_a(x)$ as well as the free propagator
\begin{equation}
 G^{(m)}_{ab}(x,x') = G_{0,ab}^{(m)} + \hat{G}^{(m)}_{ab}(x,x'),
\end{equation}
where now $\hat{G}^{(m)}_{ab}$ has the property of being finite in the IR ($m^2 \to 0$). The interaction part of the action takes the following form:
\begin{equation}
 S_{int} = \frac{\lambda V_d}{8N} |\phi_0|^4 + \tilde{S}_{int}[\phi_{0a},\hat{\phi}_a].
\end{equation}
Here $\vert\phi_0\vert^2=\phi_{0a} \phi_{0a}$ and $V_d$ is the total volume of Euclidean de Sitter
 space in $d$-dimensions, which is finite thanks to the compactification.
 
In order to compute the quantum correlation functions of the theory we define the generating functional in the presence of sources $J_{0a}$ and $\hat{J}_a$,
\begin{eqnarray}
   Z[J_0, \hat{J}] &=& \mathcal{N} \int d^N \phi_0 \int \mathcal{D}\hat{\phi} \, e^{-S - \int_x  (J_{0a} \phi_{0a} + \hat{J}_a \hat{\phi}_a )} = exp\left( -\tilde{S}_{int}\left[\frac{\delta}{\delta J_0},\frac{\delta}{\delta \hat{J}}\right] \right) Z_0[J_0] \hat{Z}_{f}[\hat{J}],
   \label{Z-int}
\end{eqnarray}
where we introduced the shorthand notation $\int_x = \int d^d x \sqrt{g}$. Here $Z_0[J_0]$ is defined as the exact generating functional of the theory with the zero modes alone, and it gives the leading IR contribution. Note that, as the zero modes are constant on the sphere, their kinetic terms vanish, and $Z_0[J_0]$ involves only ordinary integrals, which allows it to be exactly computed in several interesting cases following Ref.~\cite{Rajaraman1}. The corrections beyond the leading IR approximation come from $\tilde{S}_{int}$ which encondes the interaction among the zero and UV modes.

The effective potential gives valuable information about how the quantum fluctuations around a background field $\bar{\phi}$ influence its behavior. We are interested in particular in the generation of a dynamical mass due to quantum effects. Up to quadratic order it can be shown that 
\begin{equation}
 V_{eff} (\bar{\phi}_0) = V_0 + \frac{1}{2} \frac{N}{V_d \langle \phi_0^2 \rangle} |\bar{\phi}_0|^2 + \mathcal{O}(|\bar{\phi}_0|^4) \equiv V_0 + \frac{1}{2} m_{dyn}^2 |\bar{\phi}_0|^2 + \mathcal{O}(|\bar{\phi}_0|^4). \label{eff-pot-quad}
\end{equation}
This is an exact property of the Euclidean theory valid for all $N$ and $\lambda$, which shows that the dynamical mass $m_{dyn}$, defined by the above equation, is related to the inverse of the variance of the zero modes.

A case of great interest is when the fields are massless at tree level, $m=0$, as it is in this case in which the perturbative treatment becomes ill-defined. The nonperturbative treatment of the zero modes ensures that these modes acquire a dynamical mass, avoiding the  IR divergence associated to the free two-point functions 
in the massless limit. 
Indeed, the $n$-point functions of the zero modes can be exactly computed from $Z_0[J_0]$ to be
\begin{equation}
 \langle \phi_0^{2p} \rangle_0 = \frac{\int_0^{\infty} d\phi_0 \, \phi_0^{N-1+2p} e^{-\frac{V_d \lambda}{8N} \phi_0^4}}{\int_0^{\infty} d\phi_0 \, \phi_0^{N-1} \, e^{-\frac{V_d \lambda}{8N} \phi_0^4}}
  =2^{\frac{3p}{2}} \left( \frac{N}{V_d \lambda} \right)^{\frac{p}{2}} \frac{\Gamma\left[ \frac{N+2p}{4} \right]}{\Gamma\left[ \frac{N}{4} \right]}, \label{phi_0-2p-massless}
\end{equation}
which exhibit no IR divergences. This equation shows a scaling of the form $\phi_0 \sim \lambda^{-1/4}$, making the perturbative expansion of the UV modes to be in powers of $\sqrt{\lambda}$.

At the leading IR order the interaction between the zero and UV modes in Eq.~\eqref{Z-int} can be neglected,
\begin{equation}
 m_{dyn,0}^2 = \sqrt{\frac{N \lambda}{2V_d}} \frac{1}{2} \frac{\Gamma\left[ \frac{N}{4} \right]}{\Gamma\left[ \frac{N+2}{4} \right]}. \label{mdyn0-allN}
\end{equation}
For $N = 1$, we recover the result of \cite{Rajaraman1}, which is also the one from the stochastic approach \cite{Starobinsky}.

\section{Corrections from the UV modes to the two-point functions}

We will now show the results of computing the two-point functions of the full scalar fields including up to the second perturbative correction coming from the UV modes. Corrections to the leading order result come from expanding the exponential with $\tilde{S}_{int}$ in Eq.~\eqref{Z-int}. 

We split the  two-point functions of the total fields $\phi_a$ into UV and IR parts 
\begin{equation}
 \langle \phi_{a}(x) \phi_{b}(x') \rangle = \langle \phi_{0a} \phi_{0b} \rangle + \langle \hat{\phi}_{a}(x) \hat{\phi}_{b}(x') \rangle\, = \frac{1}{Z[0,0]} \Biggl[ \frac{ \delta^2 Z [J_0,\hat{J}] }{\delta J_{0a} \delta J_{0b}} \Bigg|_{J_0,\hat{J}=0} + \frac{ \delta^2 Z[J_0,\hat{J}] }{\delta \hat{J}_a(x) \delta \hat{J}_b(x')} \Bigg|_{J_0,\hat{J}=0} \Biggr], \label{full-2pt}
\end{equation}
where the cross-terms vanish by orthogonality.
In what follows we compute each part separately.

The expression for the UV part of the propagator can be simplified considerably using that the integrals of free UV propagators in Euclidean space can be expressed in terms of derivatives of a single propagator with respect to its mass:
\begin{equation}
\int ...\int_{x_2,..,x_{k-1}} \hat{G}^{(m)}(x_1,x_2)... \hat{G}^{(m)}(x_{k-1},x_k) = 
\frac{(-1)^k}{(k-2)!}   \frac{\partial^{k-2}\hat{G}^{(m)}(x_1,x_k)}{\partial (m^2)^{k-2}}.\label{Gk}
\end{equation} 
Using this, one can show that the  UV part reads
\begin{eqnarray}
 \langle \hat{\phi}_{a}(x) \hat{\phi}_{b}(x') \rangle = \delta_{ab} &\Biggl\{& \hat{G}^{(0)}(x,x')
 + \Biggl[ \frac{\lambda (N+2)}{2N^2} \langle \phi_0^2 \rangle_0 +\frac{\lambda}{2N}(N+2) [\hat{G}^{(0)}]_{ren} \notag \\
 &&-\frac{\lambda^2}{8N^3}(N+2)^2 V_d [\hat{G}^{(m)}]_{ren} \left( \langle \phi_0^4 \rangle_0 - \langle \phi_0^2 \rangle_0^2 \right) \Biggr] \frac{\partial \hat{G}^{(m)}(x,x')}{\partial m^2} \Bigg|_{0} \notag \\
 &&+ \frac{\lambda^2}{8N^3}(N+8) \langle \phi_0^4 \rangle_0  \frac{\partial^2 \hat{G}^{(m)}(x,x')}{\partial (m^2)^2} \Bigg|_{0} \label{2-pt-UV-ren} 
 \Biggr\}\, ,
\end{eqnarray}
while the two-point functions for the zero modes (IR part) are
\begin{eqnarray}
 \langle \phi_{0a} \phi_{0b} \rangle &=& \delta_{ab} \Biggl\{ \frac{\langle \phi_0^2 \rangle_0}{N} + \frac{\lambda}{4N^2} (N+2) \left[ \langle \phi_0^2 \rangle_0^2 - \langle \phi_0^4 \rangle_0 \right] V_d [\hat{G}^{(m)}]_{ren} \notag \\
 &&\,\,\,\,\,\,\,\,\,\,\,\,\,+ \frac{\lambda^2}{32N^3} (N+2)^2 \left[ \langle \phi_0^6 \rangle_0 - 3 \langle \phi_0^2 \rangle_0 \langle \phi_0^4 \rangle_0 + 2 \langle \phi_0^2 \rangle_0^3 \right] V_d^2 [\hat{G}^{(m)}]_{ren}^2 \notag \\
 &&\,\,\,\,\,\,\,\,\,\,\,\,\,- \frac{\lambda^2}{16N^3} (N+8) \left[ \langle \phi_0^6 \rangle_0 - \langle \phi_0^2 \rangle_0 \langle \phi_0^4 \rangle_0 \right] V_d \left( \frac{\partial [\hat{G}^{(m)}]}{\partial m^2}\right)_{fin} \Biggr\} \notag \\
 &=& \frac{\delta_{ab}}{V_d m_{dyn}^2(IR)}, \label{2-pt-IR-ren}
\end{eqnarray}
where $[...]$ indicates the coincidence limit has been taken and the $ren$ and $fin$ subscripts denote that the quantities have been rendered finite by renormalization.
The last equality follows after interpreting the corrections as a modification to the mass $m_{dyn}^2(IR)$ of the zero modes which, as mentioned before, determines the curvature of the effective potential. Eqs.~\eqref{2-pt-UV-ren} and \eqref{2-pt-IR-ren} contain the main corrections to the renormalized UV and IR propagators, for any values of $d$ and $N$. 
These are the main results of this section.

In the limit $N \to \infty$ these results are compatible with the two-point functions of the full fields, Eq.~\eqref{full-2pt}, being equal to  massive free de Sitter propagators with a dynamical mass
\begin{equation}
m_{dyn}^2 = \sqrt{\frac{\lambda}{2V_d}} + \frac{\lambda}{4} [\hat{G}^{(0)}]_{ren} + \mathcal{O}(\lambda^{3/2}).
\end{equation}
Beyond the LO contribution in $1/N$, 
the two-point functions have a more complicated structure than that of a free field.

\section{Resumming the leading IR secular terms to the two-point functions}\label{Ressum}

It is worth to note that the free UV propagators that build up the expressions of the  two-point functions of the UV modes, Eq.~\eqref{2-pt-UV-ren}, are massless. After performing the analytical continuation to the Lorentzian spacetime, this leads to an IR enhanced behavior at large distances.  Therefore,  it is necessary to extend the nonperturbative treatment to resum  the leading IR secular terms. In order to achieve this, we need to perform a resummation of diagrams that give mass to the UV propagators present in Eq.~\eqref{2-pt-UV-ren}. It can be shown that it is enough to take into account only a subclass of diagrams: those coming from the interaction term that is quadratic in both $\phi_0$ and $\hat{\phi}$,
\begin{equation}\label{biq}
 S^{(2)}_{int}[\phi_0,\hat{\phi}] = \frac{\lambda}{4N} \int d^d x \sqrt{g} \, \left( \delta_{ab} \delta_{cd} + \delta_{ac} \delta_{bd} + \delta_{ad} \delta_{bc} \right) \phi_{0a} \phi_{0b} \hat{\phi}_c \hat{\phi}_d.
\end{equation}
The remaining terms in $\tilde{S}_{int}$ are still  treated perturbatively.

We start by rewritting the generating functional Eq.~\eqref{Z-int} by grouping this term with the other terms quadratic in $\hat{\phi}$, as part of the ``free'' generating functional of the UV modes,
\begin{eqnarray}
Z[J_0, \hat{J}] &=& \mathcal{N} e^{-\tilde{\tilde{S}}_{int}\left[\frac{\delta}{\delta J_0},\frac{\delta}{\delta \hat{J}}\right]} \int d^N \phi_0 \, e^{-\left[ \frac{\lambda V_d}{8N} |\phi_0|^4 + V_d J_{0a} \phi_{0a} \right]} \notag \\
&&\times \int \mathcal{D}\hat{\phi} \, \exp\left( -\frac{1}{2} \iint_{x,y} \hat{\phi}_a  \hat{G}^{-1}_{ab}(\phi_0) \hat{\phi}_b  + \int_x \hat{J}_a \hat{\phi}_a \right)\, ,
 \label{Z-resum}
\end{eqnarray}
where now the ``free'' UV propagator $\hat{G}_{ab}(\phi_0)$ has a $\phi_0$-dependent mass,
\begin{equation}
 \hat{G}^{-1}_{ab}(\phi_0)(x,x') = \left[-\square \delta_{ab} + m_{ab}^2(\phi_0) \right] \frac{ \delta^{(d)}(x-x')}{\sqrt{g}},
 \label{inv-hatG}
\end{equation}
and where $\tilde{\tilde{S}}_{int} = \tilde{S}_{int} - S^{(2)}_{int}$ has the remaining interaction terms that are treated perturbatively.

In order to compare with the results of the previous section, it is enough to keep terms up to order $\lambda$. Therefore,
it is necessary to include perturbatively only the first correction coming from the term 
\begin{equation}
 S^{(4)}_{int}[\hat{\phi}] = \frac{\lambda}{8N} \int_x |\hat{\phi}|^4,
\end{equation}
and so the generating functional of Eq.~\eqref{Z-resum} reduces to
\begin{eqnarray}
 Z^{(1)}[J_0, \hat{J}] &=& \mathcal{N} \left[ 1 - \frac{\lambda}{8N} \delta_{cd} \delta_{ef} \int_x \frac{\delta^4}{\delta \hat{J}^4_{cdef}(x)} \right] \left( \hat{Z}_{f}\left[\hat{J}, m^2 \right] \left[ \det  \hat{G}_{rs}^{(m)} \right]^{1/2} \right)_{m\left(\frac{\delta}{\delta J_0}\right)} Z_0[J_0]. \,\,\,\,\,\,\,\,\,\,\,\, 
\end{eqnarray} 
Now we proceed to calculate the connected two-point functions 
of the UV modes,
\begin{eqnarray}
 \langle \hat{\phi}_{a}(x) \hat{\phi}_{b}(x') \rangle^{(1)} &=& \frac{1}{Z^{(1)}[0, 0]} \frac{ \delta^2 Z^{(1)}[J_0,\hat{J}] }{\delta \hat{J}_a(x) \delta \hat{J}_b(x')} \Bigg|_{J_0,\hat{J}=0} = \langle \hat{\phi}_{a}(x) \hat{\phi}_{b}(x') \rangle^{(0)} + \Delta \langle \hat{\phi}_{a}(x) \hat{\phi}_{b}(x') \rangle,
\end{eqnarray}
which we split in two contributions, according to the interaction term that we are treating perturbatively.  The calculation is technically involved, and we refer the reader
to Ref.~\cite{nos jhep} for details.
After a careful diagramatic analysis and proper renormalization
we arrive at
\begin{eqnarray}
  \langle \hat{\phi}_{a}(x) \hat{\phi}_{b}(x') \rangle^{(0)} &=& \delta_{ab} \Biggl\{ \sum_{p=0}^{\infty} \frac{1}{p!} \frac{\partial^p \hat{G}^{(m)}(x,x')}{\partial (m^2)^p} \Bigg|_{0} \left( \frac{\lambda}{2N} \right)^p \left[ 1 + \frac{(3^p-1)}{N} \right] \langle \phi_0^{2p} \rangle_0 \label{UV-prop-allN} \\
  &&- \frac{\lambda^2(N+2)}{4N} V_d [\hat{G}^{(0)}]_{ren} \times \notag \\
  && \sum_{p=0}^{\infty} \frac{1}{p!}  \frac{\partial^p \hat{G}^{(m)}(x,x')}{\partial (m^2)^p} \Bigg|_{0} \left( \frac{\lambda}{2N} \right)^p \left[ 1 + \frac{(3^p-1)}{N} \right] \left[ \langle \phi_0^{2(p+1)} \rangle_0 - \langle \phi_0^{2} \rangle_0 \langle \phi_0^{2p} \rangle_0 \right] \Biggr\}, \notag
\end{eqnarray}
and
\begin{eqnarray}
\Delta \langle \hat{\phi}_{a}(x) \hat{\phi}_{b}(x') \rangle &=& \delta_{ab} \frac{\lambda (N+2)}{2N} [\hat{G}^{(0)}]_{ren} \label{Delta_hatphi_hatphi-all-N} \\
&& \times \sum_{p=0}^{\infty} \frac{1}{p!} \frac{\partial^{p+1} \hat{G}^{(m)}(x,x')}{\partial (m^2)^{p+1}} \Bigg|_{0} \left( \frac{\lambda}{2N} \right)^p \left[ 1 + \frac{(3^p-1)}{N} \right] \langle \phi_0^{2p} \rangle. \notag
\end{eqnarray}

All of the series above can be resummed order by order in $1/N$, for which we need to expand the summands in $1/N$. Doing this, the full result for the connected two-point functions of the UV modes up to order $\lambda$ and $N^{-1}$, with a partial resummation of the infinite subset of diagrams, is:  \begin{eqnarray}
  \langle \hat{\phi}_{a}(x) \hat{\phi}_{b}(x') \rangle^{(1)} = \delta_{ab} &\Biggl\{& \hat{G}^{(m)}(x,x') + \frac{\lambda}{4} [\hat{G}^{(0)}]_{ren} \frac{\partial \hat{G}^{(m)}(x,x')}{\partial m^2} \label{UV-prop-NLO-N-1} \\
  &&+ \frac{1}{2N} \Biggl[ 2 \hat{G}^{(\sqrt{3} m)}(x,x') - 2 \hat{G}^{(m)}(x,x') \notag \\
  &&- \sqrt{\frac{\lambda}{2V_d}} \frac{\partial \hat{G}^{(m)}(x,x')}{\partial m^2} +\frac{\lambda}{2V_d} \frac{\partial^2 \hat{G}^{(m)}(x,x')}{\partial (m^2)^2} \notag \\
  &&+ \frac{\lambda}{4} [\hat{G}^{(0)}]_{ren} \left(7 \frac{\partial \hat{G}^{(m)}(x,x')}{\partial m^2} - 6 \frac{\partial \hat{G}^{(\sqrt{3} m)}(x,x')}{\partial m^2} \right) \Biggr] \Biggr\}_{m_{dyn,0}}, \notag
\end{eqnarray}  where $m_{dyn,0}^2 = \sqrt{\frac{\lambda}{2V_d}}$. Here, all the UV propagators at separated points now have a mass squared of order $\sqrt{\lambda}$. The only instance of a massless UV propagator has its coincidence limit taken, and it is therefore just a finite constant factor with no IR issues. These are our main results. It was verified as a cross-check that this expression reduces to the perturbative one of Eq.~\eqref{2-pt-UV-ren}, upon expanding the latter at NLO in $1/N$. A noteworthy observation is the presence of some propagators whose squared mass is three times that of the others, something which could not have been anticipated from the perturbative result. The large distance behaviour of the two-point functions ultimately depends on the masses of the free propagators that build up the expression, $m_{dyn,0}^2$ and $3 m_{dyn,0}^2$, which determine how fast it decays. 

One can see that the Lorentzian 2-point functions corresponding to our results Eqs.~\eqref{2-pt-IR-ren} and \eqref{UV-prop-NLO-N-1} coincide with the ones of Ref.~\cite{Gautier} when expanded up to the corresponding order, the latter given by 
\begin{equation}
 \langle \phi_a(x) \phi_b(x') \rangle = \delta_{ab}\left[ \left(1 - \frac{5}{16N} \right)\, G^{(m_+)}(x,x') + \frac{5}{16N}\, G^{(m_-)}(x,x')\right],
 \label{2-pt-Serreau}
\end{equation}
with masses $m_+^2 = m_{dyn,0}^2 \left( 1 + \frac{1}{4N} \right)$ and $m_-^2 = 5 m_+^2$. These are valid up to the NLO in the large N expansion as well, but only at the leading IR order.

\section{Conclusions}

In this work we considered an interacting $O(N)$ scalar  field model in $d$-dimensional Euclidean de Sitter space, paying particular attention to the IR problems that appear for massless and light fields. We  presented an extension of the approach of Refs.~\cite{Rajaraman1,BenekeMoch} to the $O(N)$ model. The zero modes are treated exactly while the corrections due to the interactions with the UV modes are computed perturbatively. The calculation of the two-point functions of the field shows that the exact treatment of the zero modes cures the IR divergences of the usual massless propagator: the two-point functions becomes de Sitter invariant. 

Although the massless UV propagator is de Sitter invariant, its Lorentzian counterpart exhibits a growing behavior at large distances, invalidating the perturbative expansion in this limit. This problem can be fixed in the leading order large $N$ limit by resumming the higher order corrections: one can show that the final result corresponds to  two-point functions of  free fields with a self-consistent mass.  However, the NLO contains derivatives of the free propagator of the UV modes. 
The behavior of the correlation functions in the IR limit can be improved   by performing a resummation of a class of diagrams that give mass to the UV propagator. Higher order corrections can be systematically computed in a perturbative expansion in powers of both $\sqrt\lambda$ and $1/N$. We presented explicit results up to second order in $\sqrt{\lambda}$ and NLO in $1/N$.

Our results reduce to the ones obtained by other nonperturbative approaches at leading IR order in $\sqrt{\lambda}$,  and coincides with that of the Hartree approximation in the large $N$ limit up to the second order in $\sqrt{\lambda}$.
Beyond the leading IR order, a consistent treatment of the UV sector becomes necessary.  The use of the Euclidean path integral (which is simpler than its {\it in-in} counterpart) together with the double perturbative expansion (in  $\sqrt\lambda$ and $1/N$) performed in our calculations, allowed us to further include the contribution of the UV modes. Moreover, in this framework, the precision of the calculation can be systematically improved  by computing higher order corrections. 

We are presently working on generalizing these methods to the case of negative square mass to study the restoration of symmetry due to the strong infrared effects in de Sitter space.

 \section{Acknowledgements}
 The work of FDM and LGT has been supported by CONICET and ANPCyT. LGT was also partially supported by the ICTP/IAEA Sandwich Training Educational Programme and UBA. DNL was supported by ICTP during the initial stage of this work. FDM would like to thank ICTP for hospitality during part of this work. DLN would like to thank the organizers of QUARKS 2016.

\end{document}